\documentclass[11pt,a4paper]{article}
\usepackage{amsmath,amssymb}
\usepackage{epsfig,graphicx}
\usepackage{cite}

\begin{document}

\title{\bf Reexamination of a~Bound\\ on the Dirac Neutrino Magnetic Moment\\ from the Supernova Neutrino Luminosity}
\author{A.~V.~Kuznetsov$^a$\footnote{{\bf e-mail}: avkuzn@uniyar.ac.ru},
N.~V.~Mikheev$^a$\footnote{{\bf e-mail}: mikheev@uniyar.ac.ru},
A.~A.~Okrugin$^{a}$\footnote{{\bf e-mail}: okrugin@uniyar.ac.ru}
\\
$^a$ \small{\em Yaroslavl State University} \\
\small{\em Sovietskaya 14, 150000 Yaroslavl, Russian Federation}
}
\date{}
\maketitle

\begin{abstract}
We investigate the neutrino helicity-flip process under supernova core conditions,
where the left-handed neutrinos being produced can be converted into right-handed neutrinos
sterile with respect to the weak interaction due to the interaction of magnetic moments
with plasma electrons and protons. Instead of the uniform ball model for the SN core
used in previous analyses, realistic models for radial distributions and time evolution 
of physical parameters in the supernova core are considered. We have obtained new upper limits
on the Dirac neutrino magnetic moment averaged over flavours and time from the condition
that the influence of the right-handed neutrino emission on the total cooling time scale
should be limited. 
\end{abstract}

\section{Introduction}
\label{sec:Introduction}
Nonvanishing neutrino magnetic moment leads to the helicity-flip process 
in which the left-handed neutrinos produced in the stellar interior
could convert into the right-handed ones, i.\,e., sterile with respect to the weak 
interaction. This can be important, for example, when the stellar 
energy losses are taken into account.

A considerable interest to the neutrino magnetic moment arised after 
the great event of the SN1987A in connection with the modelling of a supernova explosion,
where a gigantic neutrino outflow determines in fact the process energetics.
It means that such a microscopic neutrino characteristic as the neutrino 
magnetic moment would have a crucial influence on macroscopic properties 
of these astrophysical events. Too huge outflow of right-handed neutrinos,
produced due to the magnetic moment interaction, from the core would leave no enough energy 
to explain the observed neutrino luminosity of the supernova. 
Hence an upper bound can be established on the neutrino magnetic moment.

The neutrino helicity flip $\nu_L \to \nu_R$ under physical conditions 
corresponding to the central region of a supernova has been studied in a number of works
(see, e.\,g., Refs.~\cite{Barbieri:1988,Ayala:1999,Ayala:2000}; a more extended reference list
is given in Ref.~\cite{Kuznetsov:2007}). The process is possible due to the interaction
of the Dirac-neutrino magnetic moment with a virtual plasmon, which can be both generated and absorbed:
\begin{eqnarray}
\nu_L \to \nu_R + \gamma^*, \quad \nu_L + \gamma^* \to \nu_R \, .
\label{eq:conversion}
\end{eqnarray}

A detailed analysis of the processes~(\ref{eq:conversion}), with neutrino-helicity conversion
due to the interaction with both plasma electrons and protons via a virtual plasmon and
with taking account of polarization effects of the plasma on the photon propagator
was performed in Ref.~\cite{Kuznetsov:2007}. In particular, according to the numerical analysis
the plasma proton contribution turned out to be not only significant but even dominant. 

However, all of the previous studies~\cite{Barbieri:1988,Ayala:1999,Ayala:2000,Kuznetsov:2007} were based
on a very simplified model of the supernova core as the uniform ball with some averaged values
of physical parameters. Moreover, according to modern views, the parameter values look rather too high than typical.

The aim of this paper is to make the estimation of the Dirac neutrino magnetic moment from the limit
on the supernova core luminosity for $\nu_R$ emission by a more consistent way, taking some radial
distributions and time evolution of physical parameters from some realistic models of the supernova core.
The upper bounds are obtained on the combination of the effective magnetic
moments of the electron, muon and tau neutrinos from the condition of not-spoiling the subsequent
Kelvin\,--\,Helmholtz stage of the supernova explosion by emission of right-handed neutrinos 
during a few seconds after the collapse. 

For completeness, we consider here a general case of the magnetic moment matrix $\mu_{\nu_i \nu_j} \equiv \mu_{i j}$,
where $\nu_i, \, \nu_j$ are the neutrino mass eigenstates. For the processes with the initial electron neutrino
one should replace
\begin{eqnarray}
\mu_\nu^2 \to \mu_{\nu_e}^2 \equiv \sum\limits_i 
\left | \sum\limits_j \mu_{i j} \, U_{e j} \right |^2 \, ,
\label{eq:munu_e_eff}
\end{eqnarray}
where $U_{\ell i}$ ($\ell = e, \mu, \tau$) is the unitary leptonic mixing matrix by Pontecorvo\,--\,Maki\,--\,Nakagawa\,--\,Sakata,
and similarly for the muon and tau initial neutrinos. 
%

\section{The recent model of the O-Ne-Mg core collapse SN} 
\label{sec:Janka:2009}
The recent model was developed by H.-Th. Janka with collaborators who presented us the results
of their simulations~\cite{Janka:2009} of the O-Ne-Mg core collapse supernovae which were a continuation
of model simulations of Refs.~\cite{Kitaura:2006,Janka:2008}. The successful explosion results for this
case have recently been independently confirmed by the Arizona/Princeton
SN modelling  group~\cite{Dessart:2006,Burrows:2007} which found very similar results. So we were provided
with a model whose explosion behavior was comparatively well understood and generally accepted.

We should stress that this O-Ne-Mg core collapse model (for the initial stellar mass of $8.8 \, M_\odot$) 
is not applicable directly to SN1987A which was $15$\,--\,$20 \, M_\odot$ prior to collapse and according 
to the evolution theory it had a collapsing core which consisted of iron-peak elements. 

We integrate over the volume of the neutrino-emitting region $V$ to obtain the spectral density
of the energy luminosity of a supernova core via right-handed neutrinos: 

\begin{eqnarray}
\frac{\mathrm{d} L_{\nu_R}}{\mathrm{d} E} = 
\int \, \mathrm{d} V  \, \frac{E^3}{2 \, \pi^2} \, \Gamma_{\nu_R} (E) \, .
\label{eq:dL/dE_int}
\end{eqnarray}

\begin{figure*}[th]
\centering%
\parbox[t]{0.485\textwidth}{\centering%
\includegraphics*[width=0.480\textwidth]{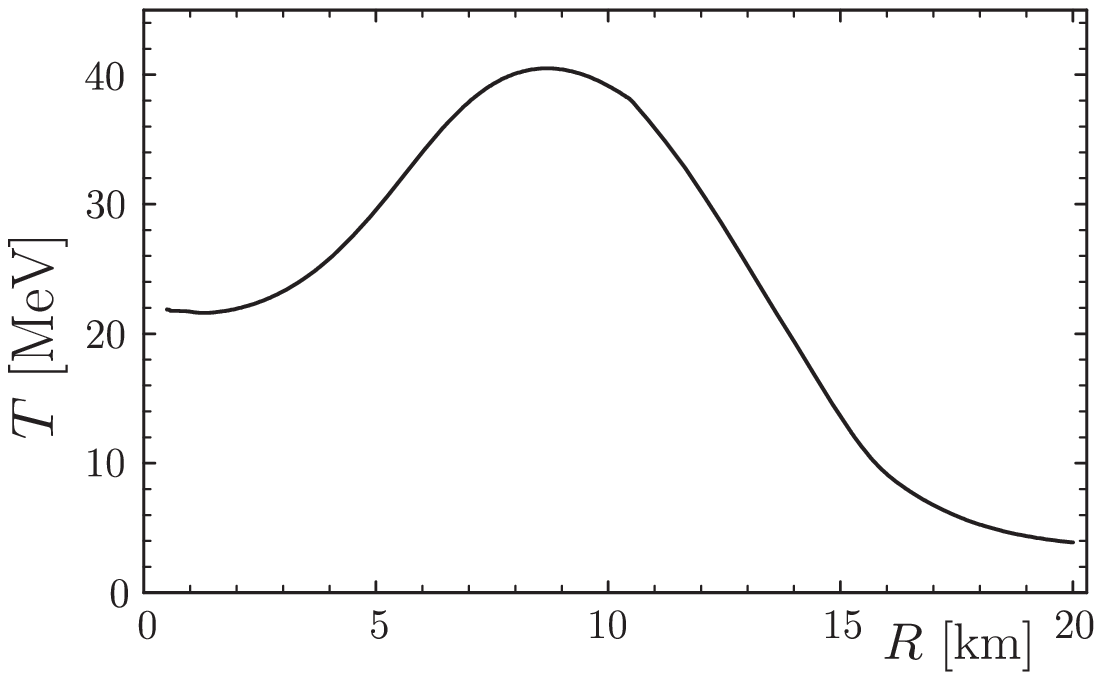}
\caption{The radial distribution for the temperature within the SN core at the moment 
$t = 1.0$ sec after the collapse~\cite{Janka:2009}.%
\label{fig:T_R}%
}}\hfil\hfil%
%
%
\parbox[t]{0.485\textwidth}{\centering%
\includegraphics*[width=0.480\textwidth]{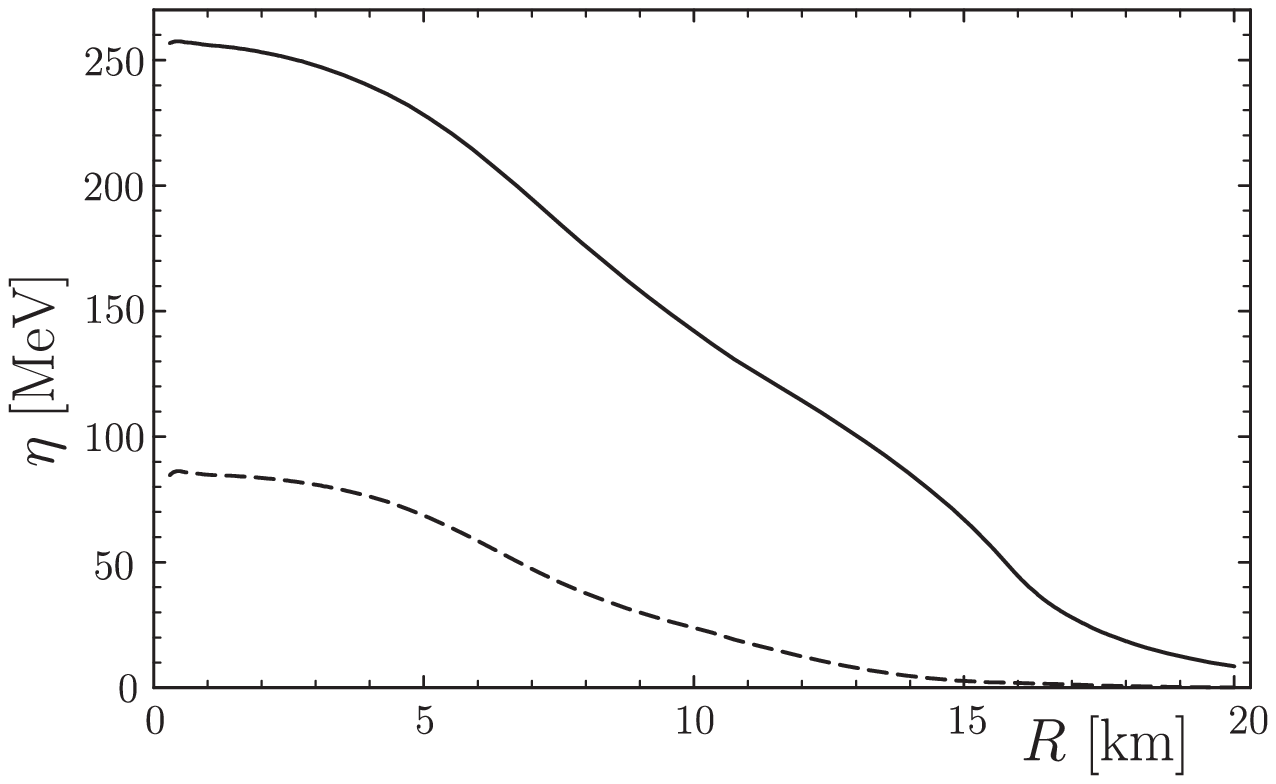}
\caption{Radial distributions for the chemical potentials of electrons (solid line)
and electron neutrinos (dashed line) within the SN core
at the moment $t = 1.0$~sec after the collapse~\cite{Janka:2009}.%
\label{fig:mue_munu_R}%
}}
%
\end{figure*}

\begin{figure}[htb]
\begin{center}
\includegraphics*[width=0.480\textwidth]{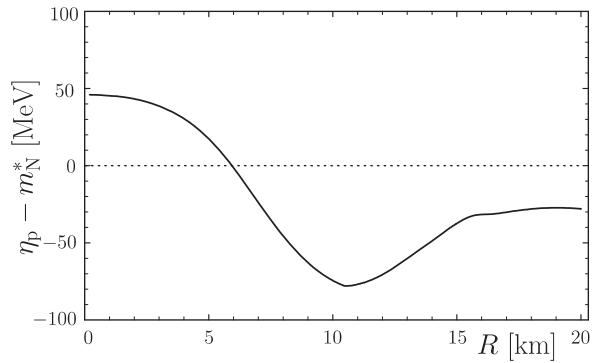}
\caption{The radial distribution for the proton nonrelativistic chemical potential
$\eta_p^* = \eta_p - m_N^*$ within the SN core at the moment $t = 1.0$ sec
after the collapse~\cite{Janka:2009}.}
\label{fig:mup_R}
\end{center}
\end{figure}

Here, taking the values defined in Eqs.~(20) and (21) and the corresponding formulas from Appendix~A
of Ref.~\cite{Kuznetsov:2007}, we take in account their dependence on the radius $R$ and time $t$.
A comprehensive set of parameter distributions used in our estimation includes
the profiles~\cite{Janka:2009} of the density $\rho$, the temperature $T$, the electron fraction
$Y_e$, the fractions of electron neutrinos $Y_{\nu_e}$, electron anti-neutrinos $Y_{\bar\nu_e}$,
and the fractions $Y_{\nu_x}$ for one kind of heavy-lepton neutrino or antineutrino ($\nu_x = \nu_{\mu, \tau},
\bar\nu_{\mu, \tau}$), which are treated identically. The time evolution of the parameter distributions
is calculated~\cite{Janka:2009} within the interval until $\sim$ $2$ sec after the collapse.
For the sake of illustration, we present in Figs.~\ref{fig:T_R}\,--\,\ref{fig:mup_R}
the radial distributions within the SN core, from $0$ to $20$ km, at the moment 
$t = 1.0$ sec after the collapse for the temperature~\cite{Janka:2009}, for the chemical potentials 
of electrons $\eta_e$ and electron neutrinos $\eta_{\nu_e}$ (calculated on the base of the data of
Ref.~\cite{Janka:2009}), and for the proton nonrelativistic chemical potential $\eta_p^* = \eta_p - m_N^*$
defining the degeneracy of protons (calculated on the base of the data of Ref.~\cite{Janka:2009} and
of the effective nucleon mass $m_N^*$ in plasma, see Ref.~\cite{Raffelt:1996}, p.~152).  

\begin{figure*}[htb]
\centering%
\parbox[t]{0.485\textwidth}{\centering%
\includegraphics*[width=0.480\textwidth]{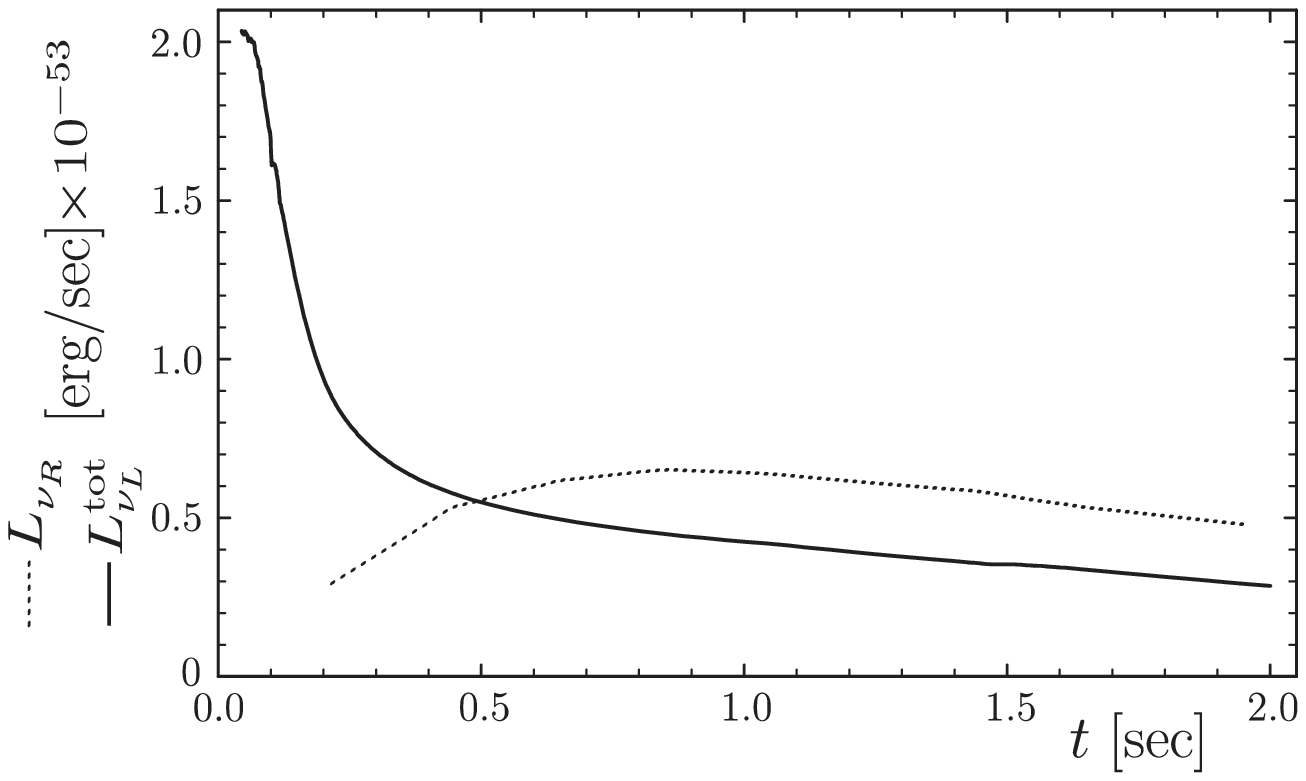}
\caption{The time evolution of the total luminosity $L_{\nu_R}^{\mathrm{tot}}(t)$ of all active neutrino species~\cite{Janka:2009} (solid line) and the right-handed neutrino luminosity $L_{\nu_R}(t)$,
Eq.~(\ref{eq:L_def}), at $\mu_\nu = 3\cdot10^{-12}\,\mu_\mathrm{B}$ (dotted line).%
\label{fig:L_nuL_t}%
}}\hfil\hfil%
%
%
\parbox[t]{0.485\textwidth}{\centering%
\includegraphics*[width=0.480\textwidth]{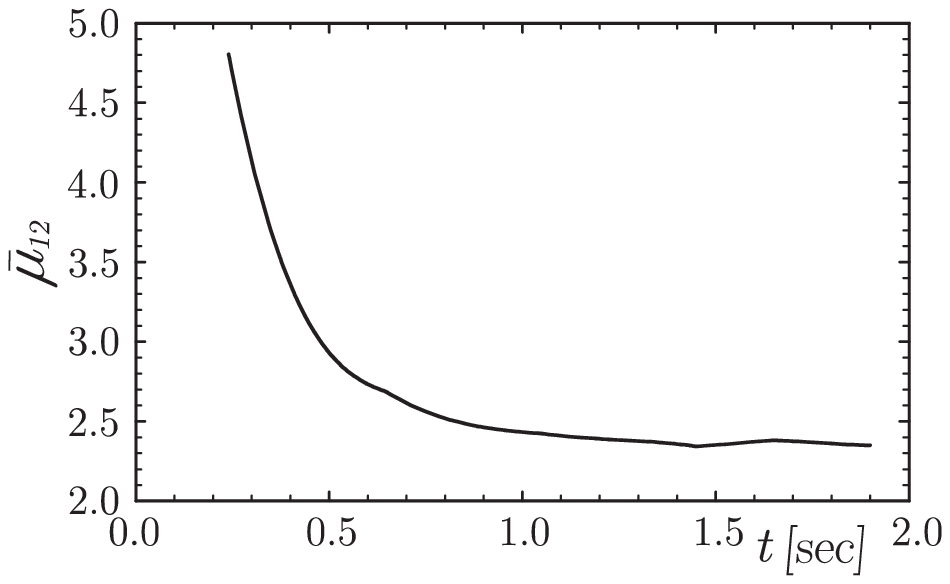}
\caption{The time evolution of the upper bound on the neutrino magnetic moment within
the time interval until $2$ sec after the collapse (in assumption that the effective magnetic moments
of electron, muon and tau neutrinos are equal).%
\label{fig:mu_nu_t}%
}}
%
\end{figure*}

To analyse the influence of the right-handed neutrino emission on the SN energy loss,
we also used the time evolution of the total luminosity of all species of left-handed 
neutrinos~\cite{Janka:2009}, presented in Fig.~\ref{fig:L_nuL_t}.

Integrating Eq.~(\ref{eq:dL/dE_int}) over the neutrino energy, one obtains 
the time evolution of the right-handed neutrino luminosity:
\begin{eqnarray}
L_{\nu_R} (t) = \frac{1}{2 \, \pi^2} \, \int \, \mathrm{d} V  
\, \int\limits_0^\infty \, \mathrm{d} E \, E^3 \, \Gamma_{\nu_R} (E) \,.
\label{eq:L_def}
\end{eqnarray}
This is a novel cooling agent which would have to compete with the energy-loss 
via active neutrino species in order to affect the total cooling time scale 
significantly. Therefore, the observed $SN1987A$ signal duration indicates that 
a novel energy-loss via right-handed neutrinos is bounded by left-handed heutrino luminosity
\begin{eqnarray}
L_{\nu_R} < L_{\nu_L} \,,
\label{eq:E_lim}
\end{eqnarray}
and we believe this estimation to be applicable also to the considered O-Ne-Mg core collapse model.
Within the considered time interval until $2$ sec after the collapse, one obtains
from Eqs.~(\ref{eq:L_def}), (\ref{eq:E_lim}) the time-dependent upper bound on the combination
of the effective magnetic moments of the electron, muon and tau neutrinos. Assuming for simplicity
that these effective magnetic moments are equal, one obtains the time evolution
of the upper bound on some flavor-averaged neutrino magnetic moment $\bar \mu_\nu$ shown
in Fig.~\ref{fig:mu_nu_t}, where $\bar \mu_{12} = \bar \mu_\nu/(10^{-12} \, \mu_{\rm B})$.

As is seen from Fig.~\ref{fig:mu_nu_t}, the averaged upper bound tends to some value, providing the limit 
\begin{eqnarray}
\bar \mu_\nu < 2.4 \, \times 10^{-12} \, \mu_{\rm B}\,.
\label{eq:mu_lim_Janka09}
\end{eqnarray}
In a general case the combined limit on the effective magnetic moments 
of the electron, muon and tau neutrinos is
\begin{eqnarray}
\left [ \mu_{\nu_e}^2 + 0.71 \left(\mu_{\nu_\mu}^2 + \mu_{\nu_\tau}^2 \right) \right]^{1/2} 
< 3.7 \, \times 10^{-12} \, \mu_{\rm B}\,,
\label{eq:mu_lim_Janka09_comb}
\end{eqnarray}
where the effective magnetic moments are defined according to Eq.~(\ref{eq:munu_e_eff}). 
This limit is less stringent than the bound~\cite{Kuznetsov:2007} obtained in the frame 
of the uniform ball model for the SN core, but it is surely more reliable. 
Additionally, the upper bound on the effective magnetic moments 
of muon and tau neutrinos is established. 

\section{Earlier models of the SN explosion} 
\label{sec:other}
The similar procedure of evaluation was performed with using of the data of 
the model~\cite{Buras:2006} by R.~Buras et al. (2006) 
of the two-dimensional hydrodynamic core-collapse 
supernova simulation for a 15 $M_\odot$ star. Namely, the radial distributions 
of parameters at the moments $t = 0.2, 0.4, 0.6, 0.8$ sec after the collapse 
in the model \textit{s15Gio\_32.a} were taken from Fig.~40 of Ref.~\cite{Buras:2006}. 
Additionally, the fraction of electron neutrinos was evaluated as 
$Y_{\nu_e} \simeq (1/5) \, Y_e$. 
Calculating the right-handed neutrino luminosity with those parameters and 
putting the limit~(\ref{eq:E_lim}), 
where the total luminosity via active neutrino species $L_{\nu_L}$ in that model 
can be taken from Fig.~42 of Ref.~\cite{Buras:2006}, 
one obtains that the upper bound on the flavor-averaged neutrino magnetic moment 
$\bar \mu_\nu$ also varies in time as in the previous case. 
The time-averaged upper bound on $\bar \mu_\nu$ corresponding 
to the interval $0.4$\,--\,$0.8$ sec, is:
\begin{eqnarray}
\bar \mu_\nu < 2.7 \, \times 10^{-12} \, \mu_{\rm B}\,,
\label{eq:mu_lim_Buras06}
\end{eqnarray}
to be compared with the limit~(\ref{eq:mu_lim_Janka09}). 

Using the results of Ref.~\cite{Pons:1999} by J.\,A.~Pons et al. (1999) where 
the thermal and chemical evolution during the Kelvin-Helmholtz phase of the birth of a
neutron star was studied, taking the data from Figs. 9 and 14, we have obtained 
the time-averaged upper bound on $\bar \mu_\nu$ for the time interval $1$\,--\,$10$ sec of 
the post-bounce evolution in the form:
\begin{eqnarray}
\bar \mu_\nu < 1.2 \, \times 10^{-12} \, \mu_{\rm B}\,.
\label{eq:mu_lim_Pons99}
\end{eqnarray}

We also used the results of Ref.~\cite{Keil:1995} by W. Keil and H.-Th. Janka (1995) where 
the numerical simulations were performed of the neutrino-driven deleptonization and 
cooling of newly formed, hot, lepton-rich neutron star. Using the data presented 
in Figs. 3\,--\,9 on the SBH model (of the hot star with a ``small'' baryonic mass), 
we have evaluated the time-averaged upper bound on $\bar \mu_\nu$ for the time interval $0.5$\,--\,$5$ sec 
after the collapse in the form:
\begin{eqnarray}
\bar \mu_\nu < 1.1 \, \times 10^{-12} \, \mu_{\rm B}\,.
\label{eq:mu_lim_Keil95}
\end{eqnarray}

One can summarize that the upper bound on the flavor- and time-averaged 
neutrino magnetic moment at the Kelvin-Helmholtz phase of the supernova 
explosion occurs to be
\begin{eqnarray}
\bar \mu_\nu < (1.1\mbox{\,--\,} 2.7) \, \times 10^{-12} \, \mu_{\rm B}\,,
\label{eq:mu_lim_summ}
\end{eqnarray}
depending on the explosion model. 

\section{Conclusion}
\label{sec:Conclusion}
The right-handed neutrino luminosity caused by the neutrino helicity-flip process 
under the conditions of the supernova core, 
where the produced left-handed neutrinos could convert due to the neutrino magnetic 
moment interaction into the right-handed neutrinos, being sterile with respect to the weak 
interaction, is reinvestigated. 
Instead of the uniform ball model for the SN core used in previous 
analyses, realistic models for radial distributions and time evolution of physical parameters in 
the SN core are considered. The upper bounds on the flavor- and time-averaged 
magnetic moment of the Dirac type neutrino are obtained in those models, 
from the condition of not-affecting the total cooling time scale 
significantly:
\begin{eqnarray}
\bar \mu_\nu < (1.1 \mbox{\,--\,} 2.7) \, \times 10^{-12} \, \mu_{\rm B}\,,
\label{eq:mu_lim_summ2}
\end{eqnarray}
depending on the explosion model. 

\section*{Acknowledgements}
\label{sec:Acknowledgements}
We thank Hans-Thomas Janka, Lorentz H\"udepohl, and Bernard M\"uller,
who provided detailed data on the radial distributions and time evolution
of physical parameters in a supernova core obtained in their model of supernova
explosion and protoneutron star cooling. We are grateful to V.\,A.~Rubakov,
G.~Raffelt, and O.\,V.~Lychkovskiy for useful remarks.

This work was performed in the framework of realization of the Federal Target Program
``Scientific and Pedagogic Personnel of Innovation Russia'' for 2009\,--\,2013
(State contract no. P2323) and was supported in part by the Ministry of Education
and Science of the Russian Federation under the Program ``Development of Scientific
Potential of Higher School'' (project no. 2.1.1/510).

%

\end{document}